\documentclass[usenatbib,onecolumn]{mnras}

\usepackage{graphicx}
\usepackage{amssymb}
\usepackage{amsmath}
\usepackage{color}
\usepackage{bm}
\usepackage{xspace}
\usepackage{pgffor}
\usepackage{upgreek}
\usepackage{nicefrac}
\usepackage[normalem]{ulem}
\usepackage[single=true]{acro}

\newcommand{\be}{\begin{equation}} 
\newcommand{\ee}{\end{equation}}

\newcommand{\ba}{\begin{eqnarray}}
\newcommand{\ea}{\end{eqnarray}}
\newcommand{\om}{\omega}

\newcommand{\curl}{{\rm curl\, }}

\newcommand\eg{\textit{e.g.,\ }}

\newcommand{\Bf}{{magnetic field}}

\newcommand{\NS}{neutron star}
\newcommand{\NSs}{{neutron stars}}

\newcommand{\EM}{electromagnetic}

\newcommand{\ms}{magnetosphere}
\newcommand{\mss}{magnetospheres}

\newcommand{\Lf}{Lorentz factor}

\def\apj{ApJ}                 
\def\apjl{ApJ}                
\def\aapr{A\&A~Rev.}          
\def\mnras{MNRAS}             
\def\nat{Nature}              

\begin{document}

\title[Frequency drifts in FRBs]{Frequency drifts in FRBs  due to  radius-to-frequency mapping in \mss\ of neutron stars}

\author[Lyutikov]{
 Maxim Lyutikov\\
 Department of Physics and Astronomy, Purdue University, West Lafayette, IN 47907-2036, USA
}


\maketitle

\begin{abstract}
We  interpret recent observations of high-to-low frequency  drifting features in the spectra of  the repeating FRBs as evidence of  sharply changing plasma properties in the emission region,  presumably the \NS\ \mss. The drifts  are then FRBs' analogues of radius-to-frequency mapping in pulsars and   Solar type-III radio burst (but not in a sense of a particular emission mechanism).  The drift rates of $\sim 100$ MHz ms$^{-1}$ at frequencies $\sim$ GHz translate to physical size of $\sim {c \om}/{\dot{\om}} \sim$ few $\times 10^8$ cm, matching the hypothesis of the FRB origin in the \mss\ of \NSs.
We suggest that reconnection events  result in  generation of upward propagating plasma beams that produce radio emission with frequency related to the decreasing local \Bf\ and plasma density.
\end{abstract}

\section{Introduction: frequency drifts in FRBs}
\label{drifts} 
Fast Radio Bursts (FRBs) \citep{2007Sci...318..777L,2019A&ARv..27....4P,2019arXiv190605878C} is a recently identified enigmatic  astrophysical  phenomenon. A particular sub-class of FRBs -  the repeating FRBs -  show similar downward drifting features in their dynamic spectra: FRB121102
\citep{2019ApJ...876L..23H}, FRB180814 \citep{2019Natur.566..235C}, and lately numerous FRBs detected by CHIME \citep{2019arXiv190803507T}. The spectrum consists of narrow frequency bands drifting with time down in frequency. The similarity in spectral behavior may hint at the physical origin of the emission.

Previously, \cite{2019arXiv190103260L} interpreted the frequency behavior in FRB121102 and FRB180814  as an indication of a some kind of a  stiff confining structures - most likely the ordered magnetic field, \eg\ in \NS\ \mss. Narrow spectral features could  then be related to the spatially  local plasma parameters (\eg\ plasma and cyclotron frequencies) or changing resonant conditions. Frequency drift then reflects the propagation of the emitting particles in changing magnetospheric conditions, similar to what is called ``radius-to-frequency mapping'' in pulsar research \citep[\eg][]{1977puls.book.....M,1992ApJ...385..282P}.

Even before  the discovery of radio emission from magnetars \cite{lyutikovradiomagnetar} \citep[see also][]{eichlerradiomagnetar} suggested that magnetar radio emission will be (is) different from pulsar emission  - it is magnetically powered in magnetars, similar to solar flares -  as opposed to rotationally powered in the case of pulsars.  \cite{lyutikovradiomagnetar}  also argued that, quote, 
``one may expect
the frequency drift of the peak of radio emission, characteristic
of type-III bursts''.  With a due uncertainty (that this prediction was not confirmed in magnetars and that not all FRBs show this behavior) we consider observations of drifting narrow band emission in FRBs as pointing to reconnection-driven  plasma processes in magnetar \mss. 

The frequency drift in FRBs is also reminiscent of type-III Solar radio bursts, whereby narrow frequency features show high-to-low temporal evolution \citep[\eg][]{1974SSRv...16..145F}.
(Type-III Solar radio bursts show temporal behavior, intensity $\propto t^{-1}$ \citep[Fig. 5 and Fig. 6 in][]{1974SSRv...16..145F}, which is similar to the observed behavior in FRBs \cite{2019ApJ...876L..23H} and \cite{2019Natur.566..235C}. We regard this similarity as a coincidence since the plasma parameters/emission mechanisms are likely to be different in the two cases.)

 Type-III Solar radio bursts arise due to the  conversion of Langmuir waves into escaping modes \citep[\eg][]{1970SvA....14..250Z}. The Langmuir waves, in turn, are excited by electron beams, presumably generated at reconnection sites. Applications of these ideas to pulsars were thoroughly investigated in the early days of pulsar research, but with no firm conclusion \citep[\eg][and reference there in]{1975ApJ...196...51R,1990MNRAS.247..529A}. (See   \cite{Melrose00Review} for critical  review of pulsar radiation theory). One of the biggest limitations in case of pulsars was a slow growth rate of Langmuir instabilities. Qualitatively, effective parallel  mass scales as $\gamma^3$ (hence it is hard to accelerate relativistic particle along the direction of its motion), and bulk relativistic motion expected on the open field lines of pulsar \mss\ further increases demands on the growth rate of  pure Langmuir instability \citep{1999JPlPh..62...65L}. 
 
 \section{Coherent radio emission from  neutrons stars }
 
 The origin of pulsar radio emission(s) remains an unsolved problem \citep[\eg][]{Melrose00Review}.  As discussed, \eg by \cite{2016MNRAS.462..941L},  three different types of coherent radio emission  from \NSs\ can be identified: (i) normal
pulses, exemplified by Crab precursor;  (ii) GPs, exemplified by Crab Main
Pulses and Interpulses; (iii)  radio emission from magnetars
(coming from the region of close field lines). Type-i and type-ii are rotationally powered, type-iii is magnetically powered.
 
 All mechanisms of coherent plasma emission involve generation of unstable particle distribution, most likely driven by a fast primary beam. 
  The most promising approach is the ``plasma maser"  \citep[\eg][]{Melrose00Review}.
  Several conditions need to be satisfied: (i) plasma supports normal modes that fall in the observed range; (ii) unstable part of particles' distribution should resonate with the normal modes; (iii) the corresponding growth rate should be fast enough; (iv) resulting modes need to escape from plasma. So far, no single suggested mechanism of \NS\ radio emission can satisfy all the criteria.
  
  The plasma maser naturally generates narrow spectral features. Narrow spectral features can appear either due to the existence of  discrete plasma normal modes (\eg related to the local \Bf\ and/or plasma frequency), see \S \ref{proper},  or due to the  resonant conditions for the emitting particles  in continuous spectrum, see \S \ref{anom}.
  
 In addition, if plasma parameters  change within the emission region, the corresponding spectral features will show spacial  evolution, that will translate into temporal changes due to time-of-flight effects. This is our basic interpretation of the drifting features in FRBs' spectra. In particular we favor a model whereby a plasma beam propagates upward in the \NS\ \ms, emitting coherent radio emission corresponding to the local (and changing) plasma parameters. 
  
  We are not in a position at this moment to qualitatively assess validity of a particular coherent emission mechanisms in FRBs \citep[\eg growth rates, see][]{2019arXiv190103260L}. Yet what we can do is scale the plasma normal modes and resonant conditions to the local plasma parameters in the \NS\ \mss\ and calculate the corresponding radius-to-frequency evolution.

  \subsection{Estimates of the size of the  emission region  from frequency drifts}

 \cite{2019ApJ...876L..23H} reported characteristic drift rate  in FRB 121102 of $ \sim  200$ MHz ms$^{- 1}$ ($\dot{\om} \approx 10^{12} $ rad s$^{- 2}$).  Associating the drift with a laterally broad (in a rotating frame)   front of emitting particle propagating with nearly  the speed of light, $d t \sim d r/c$,  in   plasma  with changing parameters, we estimate the emitting size
 \be
 d r \sim \frac{c \om}{\dot{\om}} = 3 \times 10^8 {\rm cm}
 \label{Deltar}
 \ee
 where we used $\om \sim 10^{10}$ rad $^{- 1}$ for the observed frequency. The estimate (\ref{Deltar})
 is only slightly larger than the radius of a \NS, further strengthening relation between FRBs and  \NSs. 
 (The above estimate does not take into account relativistic effects of motion along the line of sight, which generally  involve  a correction factor $\sim 1/\Gamma^2$. Such effects {\it do not}  apply to the  emission produced by different parts of  radially expanding  emission front in the rotating \ms, so that the line of sight  samples  emission regions that are not necessarily causally connected.)
  

 \subsection{Proper emission frequencies}
 \label{proper}
  
 One of the important unknown factors comes from possible relativistic motion - both due to bulk motion of the emitting plasma  with \Lf\ $\Gamma$ and due to the internal particle motion with \Lf\ $\gamma$. 
  Let us assume that the plasma in the emission region has proper (rest-frame) \Bf\ $B'$ and density $n_e'$.  Assume also that bulk motion is along \Bf, so that $B'=B$, where $B$ is the observer-frame \Bf.

 The two fundamental frequencies are then the cyclotron frequency $\om_B = e B/(m_e c)$ and the plasma  frequency $\om_p' = \sqrt{ 4 \pi  n e^2/(\gamma m_e)}$ (where we implicitly assumed that most of the motion in the plasma frame is along \Bf, so that the cyclotron frequency of the normal modes does not have a factor $\gamma$ in the denominator).  The observed frequencies are higher by a Doppler factor $\delta$. 
 
 If emission is related to the cyclotron frequency, then the observed frequency is
 \be
 \om \approx \delta \frac{ e B}{m_e c} 
 \ee
 In case of a \NS\ with surface  \Bf\  $B_{NS} =  b_q B_Q$, where $B_Q = m_e ^2 c^3/(\hbar e)$
 \be
 \om=  \delta b_q \, 511 {\rm keV} \left( \frac{r}{R_{NS}} \right)^{-3}
 \ee
 This clearly shows that radius-to-frequency mapping, $\om \propto r^{-3}$, but the corresponding frequencies are probably too high. 
 
 Next, let us consider a case when the emission is  related to the plasma frequency.
   Two parametrization of $n_e'$ with $B$ are possible:
 (i) pulsar-like, 
 normalizing the rest frame density to the GJ density \citep{GoldreichJulian}
\be
n' _{pulsar}= \kappa \frac{\Omega B}{2 \pi e c \Gamma }
\label{n11}
\ee
where $\kappa\sim 10^3-10^6$ is the observer frame multiplicity \citep{1977ApJ...217..227F,2010MNRAS.408.2092T} and $\Omega$ is the \NS\ spin;
and (ii)  magnetar-like \citep{tlk02}
\be
n' _{magnetar}=  d \phi \frac{B}{2\pi e r \Gamma}
\label{n22}
\ee
(the last comes from equating $\curl B \sim d \phi B/r$ to $(4\pi/c) 2 n e c$,  $d \phi$ is a typical twist angle in the magnetar \ms.) 

A merger of two Langmuir waves with frequency $\sim \om_p'$ in the plasma frame will produce observed radiation at \footnote{
Non-linear plasma waves conversion processes  in strong magnetic field might produce  escaping \EM\ waves \citep[\eg][]{1984Ap.....20..314M,1984Afz....20..157M}.}
\be
\om \sim \delta {\om_p'} =
\left\{ 
\begin{array}{cc}
\left(  \kappa \Omega \om_B \frac{\delta}{\gamma}\right)^{1/2} =
3 \times 10^{11} \left(  \kappa  \frac{\delta}{\gamma}\right)^{1/2}   b_q^{1/2} \left(\frac{P}{10^{-3} {\rm sec}}\right)^{-2}\left(\frac{r}{r_{LC}}\right)^{-3/2} {\rm rad \, s^{-1}},  & \mbox{for scaling (\protect\ref{n11})}\\
\left(  d \phi  \frac{\om_B c}{r}  \frac{\delta}{\gamma}\right)^{1/2} =
3 \times 10^{11}  \left( d \phi  \frac{\delta}{\gamma}\right)^{1/2}   b_q^{1/2} \left(\frac{P}{10^{-3} {\rm sec}}\right)^{-2}\left(\frac{r}{r_{LC}}\right)^{-2} {\rm rad \, s^{-1}},& \mbox{for scaling (\protect\ref{n22})}
\end{array}
\right.
\label{om1}
\ee
Both these estimates can generally produce emission at the observed radio wavelengths.

The emission frequencies (\ref{om1}) demonstrate downward frequency drift as an emitting entity propagates up in the \NS\ \ms. For dipolar \Bf\ $\propto r^{-3}$, the scalings are $\om \propto t^{-3/2}$ and $\om \propto t^{-2}$ for the two cases (assuming constant Doppler factor; for time-varying Doppler factor $t \rightarrow t/\delta(t)^2$ ).

\subsection{Evolution of the resonance condition on anomalous Doppler effect}
\label{anom}

As discussed above the local emitted frequency can be determined either by the discreetness of the plasma mode (discussed in \S \ref{proper}) or due to the evolution of the resonant condition for a continuous modes. One such possibility is the emission at  the anomalous Doppler resonance \citep{1999ApJ...512..804L,1999MNRAS.305..338L}. 
Emission is produced at
\be
\om-k_\parallel v_\parallel = - \om_B/\gamma_{res}
\ee
where $\gamma_{res}$ is the \Lf\ of fast  resonant  particles
(note the minus sign on the rhs). In strongly magnetized plasma, $\om_B \gg \om_p,  \, \om$ the dispersion is $\om/(kc)  \approx 1 - \om_p^2/( \gamma \om_B^2)$.
We expect then emission at 
\be 
\om \approx \delta \frac{\om_B^3} { \gamma_{res} \gamma \om_p^2}= \delta b_q^2 \frac{m_e^2 c^4}{\gamma_{res} \gamma \hbar^2 \Omega}
\left( \frac{r}{R_{NS}}\right)^6
 \left\{\begin{array}{cc}
 \frac{1}{\kappa} &, \mbox{for scaling (\protect\ref{n11})}\\
d \phi \frac{c}{r \Omega} &, \mbox{for scaling (\protect\ref{n22})}
\end{array}
\right.
\label{anomalous}
\ee
Eq. (\ref{anomalous}) shows very strong evolution of the emitted frequency with radius.

\section{Discussion}

We interpret drifting spectral features in FRBs as  effects of radius-to-frequency mapping within the emission region. The drift rates are consistent with the locations in \NS\ \mss.  We cannot identify a particular plasma mode/coherent emission mechanism, yet  a number of scalings with the plasma parameters are expected to produce downward frequency drifts. 

Alternative interpretations of spectral drifts involve lensing and Doppler effects. In the lensing scenarios \citep{2017ApJ...842...35C}  both upward and downward drifts are expected. Since all the FRBs show  downward drift, this make a case against lensing.
The  evolution of the frequency may also reflect  Doppler boosting - if emission is  produced at the source  at a fixed frequency due to Doppler boosting it will appear at different frequencies. In such case one then expects a strong correlation between the intensity and the peak frequency. We reject this possibility: there is no monotonic trend in the brightness of the sub-bursts, \eg Fig.  1 of  \cite{2019ApJ...876L..23H}.

Above derivations of emission properties are surely order-of-magnitude estimates, and are bound to be limited in their simplicity, as pulsar research showed for similar estimates in case of pulsars. For example, if emission is produced in \mss\ of rotating \NSs\ the Doppler factor $\delta$ will depend on time both due to rotation of the \NS\ and due to the propagation of the emitting structures in the curved \Bf\ lines.  Still, the above estimates demonstrate that there is a genuine possibility of relating the origin  of FBRs to \NS\ \mss.

Having narrowed down the most likely location of the FRBs' emission to \NSs\ \mss, there are two possible  energy sources: rotation and \Bf.  \cite{2016MNRAS.462..941L} argued that if the FRBs  are analogues of giant pulses (GPs) but coming from young (ages tens to hundreds years) pulsars with Crab-like \Bf, then the required  initial periods need to be in a few msec range -  a reasonable assumption for $D \leq$ few hundreds Mpc, but 
 the identification of the FRB host with a galaxy at $D=1$ Gpc  \citep{2017Natur.541...58C} makes this possibility unlikely \citep{2017ApJ...838L..13L}.

Our preferred model is the magnetically powered FBRs,  associated presumably   with magnetar flares \citep{2013arXiv1307.4924P}.
Let us next list a few observational  arguments for and against associating FRBs with magnetar flares. Coherent radio emission can be produced at the initial stage of a ``reconnection flare'', whereby coherent ``kinetic jets'' of particles are generated, like the ones in the studies of Crab flares \citep[\eg][]{2014PhPl...21e6501C,2017JPlPh..83f6301L,2018JPlPh..84b6301L}. But there are observational constraints: (i)  the SGR 1806 - 20 flare had peak power of $ 10^{47}$ erg s$^{-1}$  \citep{palmer} but was not seen by Parkes radio telescope \citep{2016ApJ...827...59T}; that puts an upper limit on radio-to-high -energy efficiency $\leq 10^{-6}$. For the Repeater, the first Repeater,  the implied high energy luminosity would be then $\geq 10^{47}$ erg s$^{-1}$. On the other hand,
if the  Repeater was in our Galaxy the corresponding fluxes would be in GigaJansky, which are clearly not seen. 
Also in case of PSR J1119-6127 magnetar-like X-ray bursts seem to suppress radio emission \citep{2017ApJ...849L..20A}, but this is probably related to rotationally-driven radio emission, not reconnection-driven. Thus, only some special types of magnetars  can produce FRBs.

Finally, associating FRBs with (special kinds of) magnetar flares may resolve the lack  of periodicities in the appearance of the FRBs from the Repeater(s): magnetospheric reconnection events appear randomly on closed field lines. This, combined with short, millisecond-like periods, will likely erase the signatures of the rotational period in the observed time sequence of the bursts.

In conclusion, FRB emission properties point to \mss\ of \NSs\ as the origin. Two types of mechanisms can be at work - rotationally or magnetically powered. Rotationally-powered FRB emission mechanisms \citep[\eg as analogues of Crab giant pulses][]{2016MNRAS.462..941L} are excluded by the  localization of the Repeating FRB at $\sim 1$ Gpc  \citep{2016Natur.531..202S}, as discussed by \cite{2017ApJ...838L..13L}. Magnetically-powered emission has some observational constraints, but remains theoretically viable.

\section*{Acknowledgments}
This work had been supported by DoE grant DE-SC0016369,
NASA grant 80NSSC17K0757 and  NSF grants 10001562 and 10001521. I would like to thank organizers of  SRitp  FRB workshop for hospitality, and  Jason Hessels for discussions and comments on the manuscript.  


\bibliographystyle{mnras}

  \bibliography{/Users/maxim/Home/Research/BibTex}

\end{document}